\documentclass[12pt]{article}

\usepackage{fullpage}
\usepackage{amsmath}
\usepackage[utf8]{inputenc} 
\usepackage{natbib}
\usepackage{ccicons}
\usepackage{fancyhdr}
\fancypagestyle{firstpagestyle}{
\lhead{}
\chead{}
\rhead{}
\lfoot{\parbox{0.7\textwidth}{\raggedright This work is licensed under a
  \href{http://creativecommons.org/licenses/by/3.0/deed.en_US}{Creative
    Commons Attribution 3.0 Unported License} \ccby.}}
\cfoot{}
\rfoot{\thepage}

}
\fancypagestyle{otherpagesstyle}{
\lhead{}
\chead{}
\lfoot{}
\rhead{}
\cfoot{}
\rfoot{\thepage}

}

\usepackage{hyperref}
\hypersetup{pdfborder=0 0 0}

\title{Using \texttt{R} formulae to test for main effects in the
  presence of higher-order interactions}

\author{Roger Levy}

\usepackage{Sweave}
\begin{document}

\maketitle

\thispagestyle{firstpagestyle}

\abstract{Traditional analysis of variance (ANOVA) software allows researchers
to test for the significance of main effects in the presence of
interactions without exposure to the details of how the software
encodes main effects and interactions to make these tests possible.
Now that increasing numbers of researchers are using more general regression
software, including mixed-effects models, to supplant the traditional
uses of ANOVA software, conducting such tests generally requires
greater knowledge of how to parameterize one's statistical models
appropriately.  Here I present information on how to conduct such
tests using \texttt{R}, including relevant background information and
worked examples.}

\section{Introduction}
\label{sec:introduction}

Suppose that you have predictors \texttt{X} and \texttt{Y} and
response variable \texttt{Response}, and you are interested in
performing a statistical data analysis of the effects of \texttt{X},
\texttt{Y}, and their interaction.  One thing you may be interested in
is assessing the significance of the main effect of \texttt{X}.  In
general, whenever some predictor \texttt{X} is present in a model both
as a main effect and as a part of a higher order interaction, you should be very
careful about what you mean by the ``main effect of \texttt{X}''.  
\textbf{The
key thing to remember is that under the hood, the main effect of
\texttt{X} \emph{always} really means the effect of \texttt{X} when
\texttt{Y=0}.}  If \texttt{Y} is a continuous predictor, for example,
centering it changes what the main effect of \texttt{X} means.

That being said, there are some circumstances in which it is sensible
to talk about a ``main effect of \texttt{X}'' when \texttt{X} and
\texttt{Y} interact.  If \texttt{Y} is a factor, for example, one
might want to say that there is a main effect of \texttt{X} if the
average effect of \texttt{X} across all values of \texttt{Y}
consistently goes in one direction or the other.  In these
circumstances, it's useful to know how to test for the main effect of
\texttt{X} using \texttt{R} formulae.  In certain special cases you can do it using the \texttt{aov()} function, but
sometimes you may want to do it in other cases---for example,
analyzing an experiment with crossed subject and item random effects
using \texttt{lme4}.  Thus it's useful to know the general method for
writing formuale for fitting nested models whose difference is the
presence or absence of a main effect of \texttt{X} when
\texttt{X} interacts with \texttt{Y}.

If you are just interested in how to test for main effects in the
presence of interactions, you can skip straight to
Section~\ref{sec:testing-main-effects}.
Section~\ref{sec:bit-backgr-numer} gives background detail on
\texttt{R} model formulae and contrast coding that may give you a
deeper understanding of why this works.  Section~\ref{sec:examples}
gives some examples.

\section{A bit of background on numeric predictors, categorical predictors, and interactions in \texttt{R} formulae}
\label{sec:bit-backgr-numer}

It helps to start with some review of how numeric and categorical
predictors (factors) are treated in regression modeling in \texttt{R},
and also how interactions work.  This section closely follows material
from \citet{chambers-hastie:1991statisticalModels}, the authoritative
reference.  Reviewing Section 11 of \cite{venables-etal:2013-R-intro}
is also very useful.

We start with the interpretation of interactions.  The basic way of
specifying an interaction between \texttt{X} and \texttt{Y} is
\texttt{X:Y}.  This has the following interpretation (following
\citealt{chambers-hastie:1991statisticalModels}, p.\ 25):
\begin{itemize}
\item \textbf{If \texttt{X} and \texttt{Y} are both factors} of $m$
  and $n$ levels respectively, then \texttt{X:Y} introduces an $m
  \times n$ matrix of parameters into the model, one per unique
  combination of values of \texttt{X} and \texttt{Y}.

\item \textbf{If \texttt{X} is a $n$-level factor and \texttt{Y} is
    numeric}, then \texttt{X:Y} introduces a linear effect of
  \texttt{Y} with a different slope for each level of \texttt{X}
  (hence $n$ slopes).

\item \textbf{If \texttt{X} and \texttt{Y} are both numeric}, then
  \texttt{X:Y} introduces a linear effect of the product of \texttt{X}
  and \texttt{Y} (i.e.,\ equivalent to as \texttt{I(X:Y)}, and
  introducing just one parameter).
 
\end{itemize}
Note that \texttt{X:X} is always reduced down to \texttt{X} (even for
numeric predictors).

Second comes the treatment of the \texttt{R} formula operator
\texttt{*}.  This is pretty simple:

\begin{quote}
  \texttt{X*Y} \quad is always expanded out to \quad \texttt{1 + X + Y
  + X:Y}
\end{quote}

Finally comes the question of how the inclusion of factors in one's
statistical model affects how the model is parameterized.  The
simplest way of representing the contribution of an $n$-level factor
\texttt{X} into one's statistical model is with \textsc{dummy
  coding}---introducing $n$ dummy variables with the following mapping:
\begin{quote}
\begin{tabular}{llllllllllllllll}
 Level of $X$ & $X_1$  & $X_2$ & $\dots$& $X_n$  \\
1    & 1 & 0 & $\dots$ & 0 \\
2    & 0 & 1 & $\dots$ & 0 \\
$\vdots$   & $\vdots$ & $\vdots$ & $\ddots$ & $\vdots$\\
$n$      & 0 & 0 & $\dots$ & $1$
\end{tabular}
\end{quote}
and one regression weight $\beta_i$ for each dummy variable, so that
the regression equation \verb"Response ~ X" would then be interpreted as:
\begin{gather}
  \label{eq:dummy-coding}
  \eta = \beta_1 X_1  + \beta_2 X_2 + \dots + \beta_n X_n
\end{gather}
(where $\eta$ is the linear predictor).  However, there is often
redundancy between components of model formulae. The simple formula
\verb"Response ~ X", for example, is actually interpreted with an
implicit intercept, equivalent to \verb"Response ~ 1 + X".  But adding
an extra intercept parameter into Equation~\eqref{eq:dummy-coding}
would not change the statistical model.  To eliminate this redundancy,
\texttt{R} uses \textsc{contrast coding}, which allows an $n$-level
factor to be represented with $n-1$ dummy variables.  A ``true''
contrast is one in which for each dummy variable, the sum of the
coefficients across factor levels is zero.  Such contrasts include
Helmert contrasts (which are also orthogonal) and sum or deviation
contrasts (which are not), but not the \texttt{R}-default treatment
contrasts.  \texttt{R} is fairly clever about identifying when dummy
coding cannot be used (see
\citealp{chambers-hastie:1991statisticalModels}, pp.\ 38--39, for
details).
The relevant bottom line for us here is that when true (sum-to-zero) contrasts are
used for factors, the associations of parameters correspond
intuitively to what we think of as ``main effects'' and
``interactions'': for an interaction between \texttt{X} and
\texttt{Y}, the main effect of \texttt{X} is the average (across all
levels of \texttt{Y}, not weighted by amount of data in each level)
effect of changing between levels of \texttt{X}, and likewise for the
main effect of \texttt{Y}; and  the interaction between \texttt{X} and
\texttt{Y} is the deviation of actual behavior from that predicted by
the main effects.

This puts us in a situation where we're ready to test for main effects
in the presence of interactions.  Based on the discussion immediately
above, what we want to mean by ``there is a significant main effect of
a factor \texttt{X}'' in the presence of an interaction with
\texttt{Y} is that we can conclude with some degree of statistical
certainty that the overall differences between levels of \texttt{X}
(averaged across all values of \texttt{Y}) are not all zero.
Intuitively, we want to implement that test by including the
interaction terms \texttt{X:Y} in our model, but not the main-effect
term \texttt{X}.  However, we cannot do this directly in \texttt{R}
using factors, because as we just saw, the parceling out of dummy
variables and corresponding parameters happens with respect to the
entire model formula.  So for factors (unlike for numeric variables),
\texttt{1 + X + X:Y} is the same model as \texttt{X*Y}, because
\texttt{X:Y} already introduces a separate parameter for each
combination of levels of \texttt{X} and \texttt{Y} -- the different
formulae simply lead to different parameterizations.  We can, however,
isolate the contributions of the main effect of \texttt{X} by using
true contrast coding directly to create numerical variables to
represent the different levels of \texttt{Y}.  Once \texttt{Y} is
coded with numeric variables, we can subtract out the main effect of
\texttt{X} from the full model with the interaction, because \texttt{1
  + Y + X:Y} is not equivalent to \texttt{X*Y} when \texttt{Y} is
numeric. The next section explains how to do this in practice.

\section{Testing for main effects in the presence of interactions}
\label{sec:testing-main-effects}

Assume that you are interested in testing for a main effect of
\texttt{X} in the presence of an interaction with another predictor
\texttt{Y}.  As described in the introduction, the main effect of
\texttt{X} is really just the effect of changing \texttt{X} around
while holding \texttt{Y} constant at zero.  The strategy is (i) to fit
the full model (with main effects and interactions), (ii) then fit a
reduced model with a main effect of \texttt{Y} and an interaction, but
with no main effect of \texttt{X}, and then (iii) to do a nested model
comparison between the full and reduced models to assess the
statistical significance of the main effect of \texttt{X}.  Step (ii)
is the non-trivial part, so I explain below how to do it.

\begin{enumerate}
\item 
\textbf{ If \texttt{Y} is numeric}, then you can use

  \begin{quote}
\begin{verbatim}
Response ~ X*Y - X 
\end{verbatim}
  \end{quote}

  or

  \begin{quote}
\begin{verbatim}
Response ~ Y + X:Y
\end{verbatim}
  \end{quote}

  to write the null-hypothesis model for which there's no main effect
  of X but there is an interaction.  Make sure that what you mean by
  the ``main effect of \texttt{X}'' is the effect of changing
  \texttt{X} around while \texttt{Y} stays constant at zero, because
  that is what you will get!

\item \textbf{If \texttt{Y} is a factor}, then you need to convert
  \verb"Y" into an equivalent numeric variable that will allow you to
  isolate the main effect of \texttt{X}, because when \verb"Y" is a
  factor, \verb"X*Y - X" and \verb"Y + X:Y" will be the same model
  (perhaps with a different parameterization).  By the main effect of
  \texttt{X} when \texttt{Y} is a factor, you presumably mean the
  ``across-the-board'' effect (averaging over values of \texttt{Y}) of
  changing \texttt{X}. You can get this by using \emph{any} true
  contrast scheme for \texttt{Y} (i.e., any scheme in which the
  coefficients add up do zero), because the main effect of \texttt{X}
  is the effect of changing \texttt{X} when \texttt{Y=0}, and true
  contrast schemes put ``grand average'' behavior at
  \texttt{Y=0}. Thus both \texttt{contr.sum()} and
  \texttt{contr.helmert()} work. Here's a relatively easy way to
  convert a $K$-level factor \texttt{Y} into a set of numeric
  variables consistent with a sum-coding representation:

  \begin{quote}
\begin{verbatim}
 Y.numeric <- sapply(Y,function(i) contr.sum(K)[i,])
\end{verbatim}
  \end{quote}

  If \texttt{Y} had $N$ observations, then now \texttt{Y.numeric} will
  be a $(K-1) \times N$ matrix where each row is one of the contrast
  variables.  Then you can do

  \begin{quote}
\begin{verbatim}
 Y1 <- Y.numeric[1,]
 Y2 <- Y.numeric[2,]
\end{verbatim}
  \end{quote}

and so forth up to

  \begin{quote}
\begin{verbatim}
 YKminus1 <- Y.numeric[K-1,]
\end{verbatim}
  \end{quote}

and then your null-hypothesis model becomes

  \begin{quote}
\begin{verbatim}
 Response ~ Y1 + X:Y1 + Y2 + X:Y2 + ... + YKminus1 + X:YKminus1
\end{verbatim}
  \end{quote}

In the special case where $K=2$, then \texttt{Y.numeric} winds up
simply as a vector rather than a matrix and you can enter it directly
into your regression formula:

  \begin{quote}
\begin{verbatim}
 Response ~ Y.numeric + X:Y.numeric
\end{verbatim}
  \end{quote}

\end{enumerate}

Once you have fit your reduced model, then fit the full model
\verb"Response ~ X * Y" and do a nested model comparison between the two.

\section{Examples}
\label{sec:examples}

\subsection{Simple linear regression and equivalence to \texttt{aov()}}
\label{sec:simple-line-regr}

Testing for a main effect of \texttt{X} when \texttt{X} is a two-level
factor and the model also includes a main effect of a three-level
factor \texttt{Y} and the interaction between \texttt{X} and \texttt{Y}:

\begin{Schunk}
\begin{Sinput}
> set.seed(1)
> dat <- expand.grid(X=factor(c("x1","x2")),
+                    Y=factor(c("y1","y2","y3")),
+                    repetitions=1:5)
> beta.XY <- matrix(c(1,4,5,2,3,3),2,3)
> # Create response with very small noise level so that patterns are clear
> dat$Response <- with(dat,beta.XY[cbind(X,Y)] + rnorm(nrow(dat),sd=0.1))
> with(dat,                             # empirical condition means
+      tapply(Response,list(X,Y),mean)) # closely match true effects
\end{Sinput}
\begin{Soutput}
         y1       y2       y3
x1 1.013616 5.032565 3.028430
x2 3.984901 2.011127 2.978836
\end{Soutput}
\begin{Sinput}
> Y.numeric <- sapply(dat$Y,function(i) contr.sum(3)[i,])
> dat$Y1 <- Y.numeric[1,]
> dat$Y2 <- Y.numeric[2,]
> m0 <- lm(Response ~ Y1 + X:Y1 + Y2 + X:Y2,dat)
> m1 <- lm(Response ~ X*Y,dat)
> anova(m0,m1,test="F") # Test has 1 d.f. in the numerator; no sig. main effect of X
\end{Sinput}
\begin{Soutput}
Analysis of Variance Table

Model 1: Response ~ Y1 + X:Y1 + Y2 + X:Y2
Model 2: Response ~ X * Y
  Res.Df     RSS Df Sum of Sq      F Pr(>F)
1     25 0.24372                           
2     24 0.23543  1 0.0082913 0.8452 0.3671
\end{Soutput}
\begin{Sinput}
> summary(aov(Response ~ X*Y,dat)) # Linear-regression and ANOVA results match
\end{Sinput}
\begin{Soutput}
            Df Sum Sq Mean Sq  F value Pr(>F)    
X            1   0.01   0.008    0.845  0.367    
Y            2   5.23   2.614  266.512 <2e-16 ***
X:Y          2  44.89  22.446 2288.164 <2e-16 ***
Residuals   24   0.24   0.010                    
---
Signif. codes:  0 ‘***’ 0.001 ‘**’ 0.01 ‘*’ 0.05 ‘.’ 0.1 ‘ ’ 1
\end{Soutput}
\end{Schunk}

\subsection{Mixed-effects regression}
\label{sec:mixed-effects-regr}

The same case in a mixed-effects models analysis with crossed subjects
and items, where both \texttt{X} and \texttt{Y} are within-subjects
and within-items.  Note that the null-hypothesis model \emph{includes}
random slopes for the main effect of X and Y, as well as for their
interaction (including a separate random slope for all combinations of
X and Y achieves this).

\begin{Schunk}
\begin{Sinput}
> library(mvtnorm)
> library(lme4)
\end{Sinput}
\end{Schunk}

\begin{Schunk}
\begin{Sinput}
> set.seed(1)
> M <- 12
> dat <- expand.grid(X=factor(c("x1","x2")),Y=factor(c("y1","y2","y3")),
+                    subj=factor(paste("S",1:M)),item=factor(paste("I",1:M)))
> dat$XY <- with(dat,factor(paste(X,Y)))
> beta.XY <- matrix(c(1,4,5,2,3,3),2,3)
> b.S <- rmvnorm(M,rep(0,6),diag(6)/100)
> b.I <- rmvnorm(M,rep(0,6),diag(6)/100)
> dat$Response <- with(dat,beta.XY[cbind(X,Y)] + 
+                      b.S[cbind(subj,XY)] + 
+                      b.I[cbind(item,XY)] + 
+                      rnorm(nrow(dat),sd=0.1))
> Y.numeric <- sapply(dat$Y,function(i) contr.sum(3)[i,])
> dat$Y1 <- Y.numeric[1,]
> dat$Y2 <- Y.numeric[2,]
> m0 <- lmer(Response ~ Y1 + X:Y1 + Y2 + X:Y2 + (XY|subj) + (XY|item),dat,REML=F)
> m1 <- lmer(Response ~ X*Y + (XY|subj) + (XY|item),dat,REML=F)
\end{Sinput}
\end{Schunk}

\begin{Schunk}
\begin{Sinput}
> anova(m0,m1)
\end{Sinput}
\begin{Soutput}
Data: dat
Models:
m0: Response ~ Y1 + X:Y1 + Y2 + X:Y2 + (XY | subj) + (XY | item)
m1: Response ~ X * Y + (XY | subj) + (XY | item)
   Df     AIC     BIC logLik deviance  Chisq Chi Df Pr(>Chisq)
m0 48 -1040.6 -812.01 568.28  -1136.6                         
m1 49 -1038.7 -805.35 568.33  -1136.7 0.0999      1      0.752
\end{Soutput}
\end{Schunk}

If I were to report this mixed-effects regression analysis in a
journal article I might say:
\begin{quote}
  We tested for a main effect of \texttt{X} by converting \texttt{Y}
  to a sum-coding numeric representation and conducting a
  likelihood-ratio test between mixed-effects models differing only in
  the presence or absence of a fixed main effect of \texttt{X}.  Both
  models included in their fixed effects an intercept, a main effect
  of \texttt{Y}, and an interaction between \texttt{X} and \texttt{Y}.
  Both models also had maximal random effects structure, namely random
  intercepts plus random slopes for \texttt{X}, \texttt{Y}, and their
  interaction for both subjects and items. The likelihood-ratio test
  showed no evidence for a main effect of \texttt{X}
  ($p=1$, 1 d.f.).
\end{quote}

\section*{Acknowledgments}

Special thanks to Kevin Tang for pointing out a minor error in
Section~\ref{sec:testing-main-effects} of \texttt{v1} of this paper.

\bibliographystyle{apalike}
\bibliography{rpl-journals-long,rpl}

\end{document}